\documentclass[prl,showpacs,superscriptaddress,twocolumn,preprintnumbers,amsmath,amssymb,floatfix,a4paper]{revtex4-1}

\usepackage{graphicx}

\usepackage{dcolumn}
\usepackage{bm}
\usepackage{amsmath}

\graphicspath{{graphics/}}  
\usepackage{color}
\usepackage{cancel}
\usepackage{ulem}

\begin{document}


\title{Brillouin light scattering by magnetic quasi-vortices in cavity optomagnonics}

\author{A.~Osada}
\email{alto@iis.u-tokyo.ac.jp}
\email{Current affiliation: Institute for Nano Quantum Information Electronics, The University of Tokyo, Meguro-ku, Tokyo 153-8505, Japan.}
\author{A.~Gloppe}
\author{R.~Hisatomi}
\author{A.~Noguchi}
\author{R.~Yamazaki}
\affiliation{Research Center for Advanced Science and Technology (RCAST), The University of Tokyo, Meguro-ku, Tokyo 153-8904, Japan}
\author{M.~Nomura}
\affiliation{Institute of Industrial Science (IIS), The University of Tokyo, Meguro-ku, Tokyo 153-8505, Japan}
\author{Y.~Nakamura}
\affiliation{Research Center for Advanced Science and Technology (RCAST), The University of Tokyo, Meguro-ku, Tokyo 153-8904, Japan}
\affiliation{Center for Emergent Matter Science (CEMS), RIKEN, Wako, Saitama 351-0198, Japan}
\author{K.~Usami}
\email{usami@qc.rcast.u-tokyo.ac.jp}
\affiliation{Research Center for Advanced Science and Technology (RCAST), The University of Tokyo, Meguro-ku, Tokyo 153-8904, Japan}

\date{\today}
%
\begin{abstract}
A ferromagnetic sphere can support \textit{optical vortices} in forms of whispering gallery modes and \textit{magnetic quasi-vortices} in forms of magnetostatic modes with non-trivial spin textures. These vortices can be characterized by their orbital angular momenta. We experimentally investigate Brillouin scattering of photons in the whispering gallery modes by magnons in the magnetostatic modes, zeroing in on the exchange of the orbital angular momenta between the optical vortices and the magnetic quasi-vortices. We find that the conservation of the orbital angular momentum results in different nonreciprocal behaviors in the Brillouin light scattering.  New avenues for chiral optics and opto-spintronics can be opened up by taking the orbital angular momenta as a new degree of freedom for cavity optomagnonics.

\begin{description}
\item[PACS numbers]
\end{description}
\end{abstract}

\maketitle
Vortex-like excitations, which can be characterized by a topological charge, or an orbital angular momentum (OAM), have been intensively studied in the past decades and are still a hot topic in both scientific and technological context.  Optical vortex~\cite{Allen}, for instance, is utilized for manipulating microparticles~\cite{Grier} and atoms~\cite{Kozuma, Phillips}, for OAM multiplexing in telecommunication~\cite{Wang}, and encoding quantum information~\cite{Zeilinger, Inoue}. In the condensed matter research, spin textures in spin systems, that is, magnetic vortices, have been and still are a focus in the context of magnetic domain walls~\cite{Slonczewski} as well as magnetic skyrmions~\cite{NT}, e.g., in the scope of using them for information storage.

Interaction between optical vortices and magnetic vortices is expected to show novel physics owing to the exchange of the OAM. A system providing a platform for such an interaction is a ferromagnetic sphere, supporting both optical whispering gallery modes (WGMs)~\cite{Vahala} for photons and magnetostatic (Walker) modes~\cite{Walker, FletcherBell} for magnons.  We refer to the WGMs as \textit{optical vortices} and the Walker modes as \textit{magnetic quasi-vortices}, where the prefix ``quasi-" emphasizes the facts that 1) magnon is a quasi-particle with a finite lifetime and 2) OAM for the Walker mode are approximately well-defined because of the axial-symmetry-broken dipolar interaction~\cite{PRB}.

 An emerging field called \textit{cavity optomagnonics} aiming at enhancing magnon-induced Brillouin light scattering~\cite{Haigh1, AO1, Tang1, Haigh2, Kusmin, Liu, SBB} takes place in such a platform. Up to now, the magnetostatic modes used in the experiments of cavity optomagnonics~\cite{Haigh1, AO1, Tang1, Haigh2} remain limited to the uniform precession mode, or the Kittel mode, having no OAM, which has also been utilized for experiments on the \textit{circuit} quantum magnonics~\cite{Tabuchi1, Kostylev, ZhangTang, Hu, Tabuchi2, Dany} as well as for the coherent conversion between microwave and optical photons~\cite{Hisatomi}.

\begin{figure}[t]
\includegraphics[width=8.6cm]{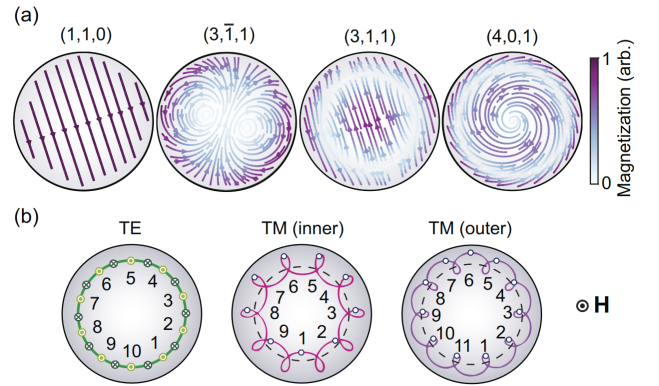}
\caption{\label{FigTexture} (a) Transverse magnetization distributions of $(1, 1, 0)$, $(3, \bar{1}, 1)$, $(3, 1, 1)$ and $(4, 0, 1)$ Walker modes in the equatorial plane.  (b) Schematic representations of the heads of the electric fields of TE and TM WGMs on the equatorial plane.  These illustrate the difference of OAM for the TE WGM, the inner component of TM WGM and the outer component of the TM WGM for the case of $m = 10$.  The hollow circles locate points where the field vectors direct upward for the ease of counting the numbers of the phase rotations in Cartesian coordinates.}
\end{figure}

In this Letter, the Brillouin light scattering hosted in a ferromagnetic sphere is experimentally investigated with a special emphasis on the magnetostatic modes involving magnetic quasi-vortices. The experiments reveal that the scattering is either nonreciprocal or reciprocal depending on the OAM of the magnons. We analyze the results with the theory developed in the accompanying paper~\cite{PRB} and find reasonable agreements. The OAM can thus be considered as a new degree of freedom in cavity optomagnonics, providing a new approach to chiral quantum optics~\cite{Lodahl2017} and topological photonics~\cite{Taylor2014,Soljacic2014}.

Let us first describe the characteristics of the Walker modes and the WGMs in terms of OAM. Henceforth, we shall suppose that the symmetry axis is $z$-axis along which a static magnetic field is applied, perpendicular to the plane of the WGM orbit. The Walker modes are nominally labeled by three indices $(n, m_{\mathrm{mag}}, r)$~\cite{Walker, FletcherBell}, where the second index $m_{\mathrm{mag}}$ is most relevant and related to the OAM $\mathcal{L}_{z}^{(m_{\mathrm{mag}})}$ as $\mathcal{L}_{z}^{(m_{\mathrm{mag}})} \simeq -\left( m_{\mathrm{mag}}-1 \right)$ under proper approximations~\cite{PRB}, while $n$ and $r$ are related to the profiles along polar and radial directions, respectively. 

In Fig.~\ref{FigTexture}(a), the distributions of the magnetization component perpendicular to the symmetry axis for some Walker modes are exemplified~\cite{PRB, Walker, FletcherBell}. For these four modes labeled by $(1, 1, 0)$,  $(3, \bar{1}, 1)$, $(3, 1, 1)$, and $(4, 0, 1)$, the OAM are $\mathcal{L}_{z}^{(1)}=0$, $\mathcal{L}_{z}^{(-1)}=2$, $\mathcal{L}_{z}^{(1)}=0$, and $\mathcal{L}_{z}^{(0)}=1$, respectively.  It is apparent that the modes with non-zero OAM exhibit topologically non-trivial spin textures, that is, \textit{magnetic quasi-vortices}.

Next we consider the OAM of the WGM. Let the optical field orbiting counterclockwise (CCW) be decomposed into three components based on their spins along the symmetry axis $z$. A transverse-electric (TE) WGM, whose direction of the field is parallel to $z$-axis and schematically shown in the left panel of Fig.~\ref{FigTexture}(b), is $\pi$-polarized.  If the TE WGM has the azimuthal mode index of $m_{\mathrm{TE}}$, the light field oscillates $m_{\mathrm{TE}}$ times in the CCW round trip and therefore the OAM reads as $\mathcal{L}_{z}^{(\mathrm{CCW, TE}, m_{\mathrm{TE}})} = m_{\mathrm{TE}}$. On the other hand, a transverse-magnetic (TM) WGM has the polarization in the plane of the WGM orbit which is decomposed into $\sigma^+$ and $\sigma^-$ components. The OAM are different for these two components. The situation is shown in the center and right panels of Fig.~\ref{FigTexture}(b). The trajectories of the head of the polarization vector for each electric field are shown. For the ease of counting the numbers of rotations of the head, hollow circles indicate positions around which the electric field directs upward. When the mode index is 10, the orbital angular momentum reads 10, 9 and 11 for the TE, inner TM, and outer TM components, respectively~\cite{PRB}. Here we call the polarization component giving the smaller (larger) number of phase rotation the \textit{inner} (\textit{outer}) component, since it actually comes from the inner (outer) region of the WGM distribution as a result of the spin-Hall effect of light~\cite{Murakami2, Bliokh1}. The origin of such an effect is the transversality of electromagnetic field which couples spin (polarization) and orbital degrees of freedom of light~\cite{Bliokh2}. Consequently, the OAM is not a good quantum number for the TM WGM and given by $\mathcal{L}_{z}^{(\mathrm{CCW, TM+}, m_{\mathrm{TM}})} = m_{\mathrm{TM}}-1$ for the inner component and $\mathcal{L}_{z}^{(\mathrm{CCW, TM-}, m_{\mathrm{TM}})} = m_{\mathrm{TM}}+1$ for the outer component, respectively. Here the ``TM$\pm$" in the superscript refers to the OAM accompanying the $\sigma^{\pm}$-polarized component. Note that the total angular momentum defined by the sum of the OAM and spin angular momentum for each of these two components is the same and be a good quantum number.

For the clockwise (CW) orbit, on the other hand, the OAM of these inner and outer components of the TM WGM are given with the proper sign inversion, that is, $\mathcal{L}_{z}^{(\mathrm{CW, TM+}, m_{\mathrm{TM}})} = -\left( m_{\mathrm{TM}}+1 \right)$ and $\mathcal{L}_{z}^{(\mathrm{CW, TM-}, m_{\mathrm{TM}})} = -\left( m_{\mathrm{TM}}-1 \right)$. Note that the OAM of the TE WGM orbiting clockwise (CW) is straightforwardly given by $\mathcal{L}_{z}^{(\mathrm{CW, TE}, m_{\mathrm{TE}})} = -m_{\mathrm{TE}}$.  

\begin{figure}[t]
\includegraphics[width=8.6cm]{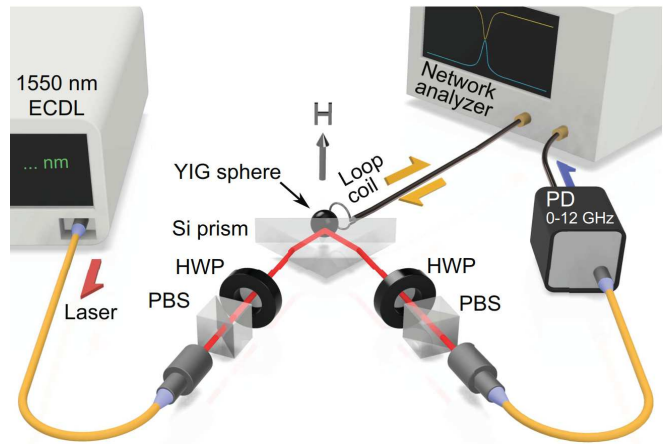}
\caption{\label{Fig1} Experimental setup for observing the Brillouin scattering.  External-cavity diode laser (ECDL) supplies the laser light of $1550$-nm wavelength. Its polarization is defined by a polarization beam splitter (PBS) and a half-wave plate (HWP) to be coupled to a particular family of the YIG sphere WGM by the intercession of the silicon prism. Microwaves from a network analyzer excite magnons via a loop coil.  Scattered and unscattered light interfere through a HWP and a PBS, and the beat signal is detected by a high-speed photodetector (PD) and analyzed by the network analyzer.}
\end{figure}

\begin{figure}[t]
\includegraphics[width=8.6cm]{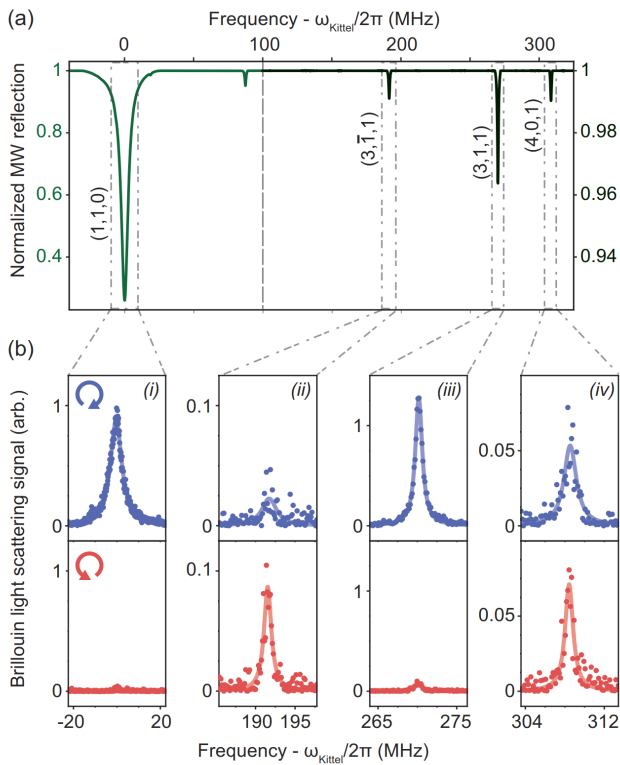}
\caption{\label{FigS11andS21} (a) Normalized microwave reflection spectrum with the mode indices assigned to the Walker modes of our interest.  Along the horizontal axis, the relative frequency to the Kittel-mode frequency $\omega_{\mathrm{Kittel}}/2\pi$ is used.  Note that the spectra below (green, left axis) and above (black, right axis) $100$~MHz are separately scaled.   (b) The signals of the Brillouin scattering of the CW (blue) and the CCW (red) WGMs by (i) $(1, 1, 0)$, (ii) $(3, \bar{1}, 1)$, (iii) $(3, 1, 1)$ and (iv) $(4, 0, 1)$ Walker modes.  Solid lines are Lorentzian fittings to the data as a guide to the eye.}
\end{figure}

We experimentally investigate how these OAM of the magnons and the photons are exchanged in a ferromagnetic sphere. The sample under consideration is a yttrium iron garnet (YIG) sphere with a diameter of $1$~mm. We focus on the Brillouin scattering induced by the magnetic quasi-vortices with $\mathcal{L}_z^{(m_{\mathrm{mag}})}=0, 1, 2$, which include the  $(1, 1, 0)$, $(3, 1, 1)$, $(4, 0, 1)$ and $(3, \bar{1}, 1)$ Walker modes. 

The experimental setup is schematically shown in Fig.~\ref{Fig1}. Laser light with a wavelength of $1550\,$nm is coupled to WGMs in the YIG sphere via a silicon prism.  The quality factor of WGMs in the YIG sphere is around $10^5$.  Whether the TE or the TM family of WGMs is coupled depends on the linear polarization before the prism, which is determined by a polarization beam splitter (PBS) and a half-wave plate (HWP). The input polarization and the frequency of the laser light is tuned to couple into a certain TM WGM on resonance. Static magnetic field around $0.25\,$T, giving a Kittel mode frequency of  $7.1\,$GHz, saturates the magnetization of the sample in the direction perpendicular to the plane of the WGM orbit. Microwaves are sent from a network analyzer to a loop coil and excite magnons in a specific Walker mode. The photons in the TM WGM are then scattered simultaneously creating (annihilating) a magnon and generating a Stokes-sideband (anti-Stokes-sideband) photon into the TE WGM~\cite{PRB}. A PBS and a HWP after the prism make the scattered and the unscattered photons interfere. The beat signal at the magnon frequency is detected by a high-speed photodetector (PD) and sent back to the network analyzer.  

The microwave reflection spectrum is plotted in Fig.~\ref{FigS11andS21}(a) with the indices of the identified Walker modes next to the observed dips~\cite{Supp}. Spectra in Fig.~\ref{FigS11andS21}(b) are the observed signals of the scattered light generated by each of the four Walker modes for CW (blue) and CCW orbits (red). For the Walker modes possessing no OAM, that is, $(1, 1, 0)$ and $(3, 1, 1)$ modes, the Brillouin scattering tends to occur more prominently for the CW orbit than that for the CCW orbit. As for $(4, 0, 1)$ mode, which has the orbital angular momentum $\mathcal{L}_{z}^{(0)}=1$, the signal strengths of the Brillouin scattering for the CW orbit and that for the CCW orbit are more or less equal.  As for $(3, \bar{1}, 1)$ mode where the OAM reads $\mathcal{L}_{z}^{(-1)}=2$, however, the Brillouin scattering tends to occur more prominently for the CCW orbit than for the CW orbit. 

\begin{figure}[t]
\includegraphics[width=8.6cm]{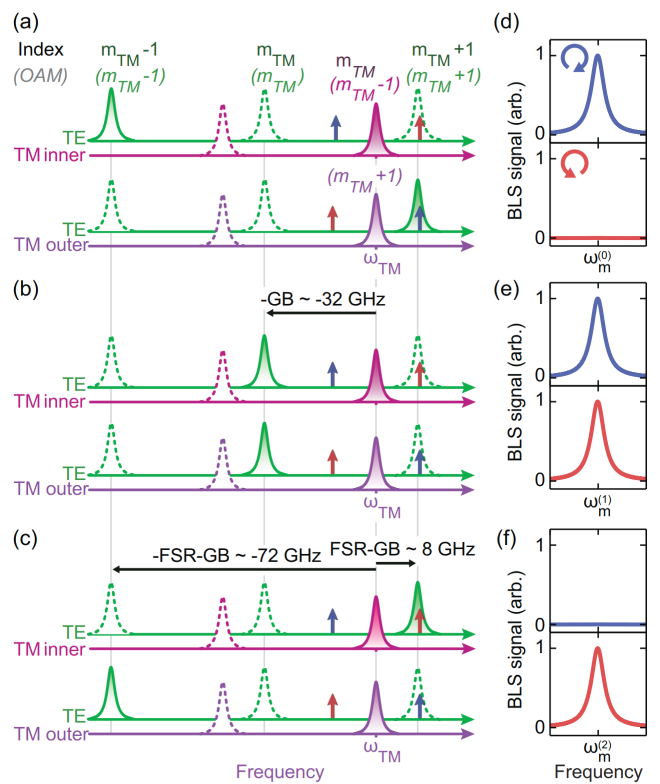}
\caption{\label{FigAnalysis}(a)-(c) Densities of states of the relevant WGMs (highlighted, otherwise dotted) for the magnon-induced Brillouin light scattering.  Here we assume that the input mode is TM with an azimuthal index of $m_\mathrm{TM}$ at a frequency of $\omega_\mathrm{TM}/2\pi$. The TM modes are split into inner (pink) and outer (purple) components, with the OAM of $m_\mathrm{TM}-1$ and $m_\mathrm{TM}+1$, respectively. Due to the geometric birefringence (GB), in our experiment the TE modes (green) are shifted by $\sim-32\,$GHz in comparison to the TM modes. Two adjacent modes of the same polarization are separated by the free spectral range (FSR) $\sim40
\,$GHz.  (a), (b) and (c) depict the situations for the considered Walker-mode OAM of $\mathcal{L}_{z}^{(1)}=0$, $\mathcal{L}_{z}^{(0)}=1$ and $\mathcal{L}_{z}^{(-1)}=2$, respectively. The scattered photons are depicted by blue and red arrows, representing respectively clockwise (CW) and counterclockwise (CCW) input orbits. (d)-(f): Theoretically predicted nonreciprocal/reciprocal behavior of the Brillouin light scattering (BLS) for the CW- and the CCW-orbit inputs, respectively for the cases (a)-(c).  The symbol $\omega_\mathrm{m}^{(i)}$ represents the Walker-mode frequency where $i$ designates the OAM.}
\end{figure}
These results can be explained by the following selection rule in the Brillouin light scattering process~\cite{PRB}. First, the light is supposed to be injected into the TM WGM with the mode index $m_{\mathrm{TM}}$ orbiting CCW.   According to the selection rule, the photons in the TM WGM would be scattered into the TE WGM with the mode index $m_{\mathrm{TE}}$, which must satisfy
\begin{equation}
m_{\mathrm{TE}} = m_{\mathrm{TM}}-m_{\mathrm{mag}} \label{eq:CCW_a}
\end{equation}
in the anti-Stokes scattering process associated with the inner component of the input TM WGM, and \begin{equation}
m_{\mathrm{TE}} = m_{\mathrm{TM}}+m_{\mathrm{mag}} \label{eq:CCW_s}
\end{equation}
in the Stokes scattering process associated with the outer component of the input TM WGM. Contrariwise, if the light is injected into the CW-orbiting TM WGM, the selection rules become
\begin{equation}
m_{\mathrm{TE}} = m_{\mathrm{TM}}+m_{\mathrm{mag}} \label{eq:CW_a}
\end{equation}
in the anti-Stokes scattering process, which is now associated with the outer component of the input TM WGM, and 
\begin{equation}
m_{\mathrm{TE}} = m_{\mathrm{TM}}-m_{\mathrm{mag}} \label{eq:CW_s}
\end{equation}
in the Stokes scattering process associated with the inner component of the input TM WGM.

Now, let us see how the selection rule given above and the spectral properties of the WGMs dictates the Brillouin scattering process. Figures~\ref{FigAnalysis}(a)-(c) show schematically the densities of states of the TM (purple) and the TE (green) WGMs. Here, for a given azimuthal mode index $m_{\mathrm{TM}}$ the resonance frequencies of the TM and the TE WGMs are different due to the geometric birefringence (GB)~\cite{SB, Lam, Schiller} by $\sim 32$~GHz, which is adopted from the experimental value, as well as the free spectral range (FSR) of $\sim 40$~GHz.  Two sets of the WGMs are depicted (top and bottom), one for the inner component [$\sigma_{+}\ (\sigma_{-})$ component for the CCW (CW) orbit] and the other for the outer component [$\sigma_{-}\ (\sigma_{+})$ component for the CCW (CW) orbit].

First, we consider the Walker modes with the OAM $\mathcal{L}_{z}^{(1)}=0$, that is, $(1, 1, 0)$ and $(3, 1, 1)$ modes.  In this case, we have the selection rules Eqs.~(\ref{eq:CCW_a})--(\ref{eq:CW_s}) with $m_{\mathrm{mag}}=1$. The relevant TE WGMs specified by the selection rule are highlighted (otherwise dotted) in Fig.~\ref{FigAnalysis}(a). In Fig.~\ref{FigAnalysis}(a), the expected frequencies of the TE photons to be created are indicated by blue and red upright arrows for the CW and the CCW orbits, respectively. We see that whether the Stokes scattering or the anti-Stokes scattering occurs out of the inner and the outer components of the TM WGM depends on the direction of the WGM orbit. As for the inner component (pink), the scattered light for both the CW and the CCW orbits are far detuned from the relevant TE WGM specified by the selection rule by $\sim \mathrm{FSR}+\mathrm{GB} = 72$~GHz~[cf. the arrow in (c)].  On the other hand, for the outer component (purple) the scattered light is almost resonant for the CW orbit and off-resonant for the CCW orbit. In Fig.~\ref{FigAnalysis}(d), the theoretically predicted nonreciprocal behavior of the Brillouin scattering is schematically shown, where the blue and red curves are respectively for the CW and the CCW cases.  This is in good agreement with the experimental results shown in Figs.~\ref{FigS11andS21}~(b, i) and (b, iii).

The same argument applies for other cases. In Figs.~\ref{FigAnalysis}(b) and (c), the relevant TE WGMs specified by the selection rules are highlighted (otherwise dotted) as in Fig.~\ref{FigAnalysis}(a). For $(4, 0, 1)$ mode with the OAM $\mathcal{L}_{z}^{(0)}=1$, all the selection rule becomes simply $m_{\mathrm{TE}} = m_{\mathrm{TM}}$ resulting in the absence of the nonreciprocity as indicated in Fig.~\ref{FigAnalysis}(e). The observed reciprocity in Fig.~\ref{FigS11andS21}(b, iv) can thus be explained. As for the Walker mode with the OAM $\mathcal{L}_{z}^{(-1)}=2$ such as $(3, \bar{1}, 1)$ mode, the selection rules are exactly the same as for the modes with the OAM $\mathcal{L}_{z}^{(1)}=0$ but the roles of the CCW and the CW orbits are interchanged. This leads to the prediction of the nonreciprocal behavior depicted in Fig.~\ref{FigAnalysis}(f), which is again in good agreement with the experimental result shown in Fig.~\ref{FigS11andS21}(b, ii).

In summary, we experimentally investigated the Brillouin light scattering within a ferromagnetic sphere and observed nontrivial nonreciprocal/reciprocal behavior of the Brillouin light scattering by the magnons with and without orbital angular momenta.
We showed that the observed phenomena can be understood in terms of conservative exchange of the orbital angular momentum between the photons in whispering gallery modes and the magnons in the Walker mode. Nonreciprocity and topology are at the heart of the current research activities in photonics, spintronics, and magnonics. Our findings suggest that cavity optomagnonics embrace these two concepts and could thus be an alternative and attractive platform for further developing chiral and topological devices.

We are grateful to Y. Shikano, Y. Tabuchi, Y. Nakata, S. Ishino, G. Tatara, S. Iwamoto, J. Haigh, S. Sharma, Y. M. Blanter, and G. E. W. Bauer for fruitful discussions. This work was supported by JSPS KAKENHI (Grant No. 15H05461, No.~16F16364, No. 26220601, and No. 26600071), Murata Science foundation, Inamori Foundation, and JST ERATO project (Grant No. JPMJER1601).  AG is an Overseas researcher under Postdoctoral Fellowship of JSPS.

%
%
%

\newpage

\section*{Supplemental Material: Brillouin light scattering by magnetic quasi-vortices in cavity optomagnonics}

\begin{figure*}
\includegraphics[width = 15cm]{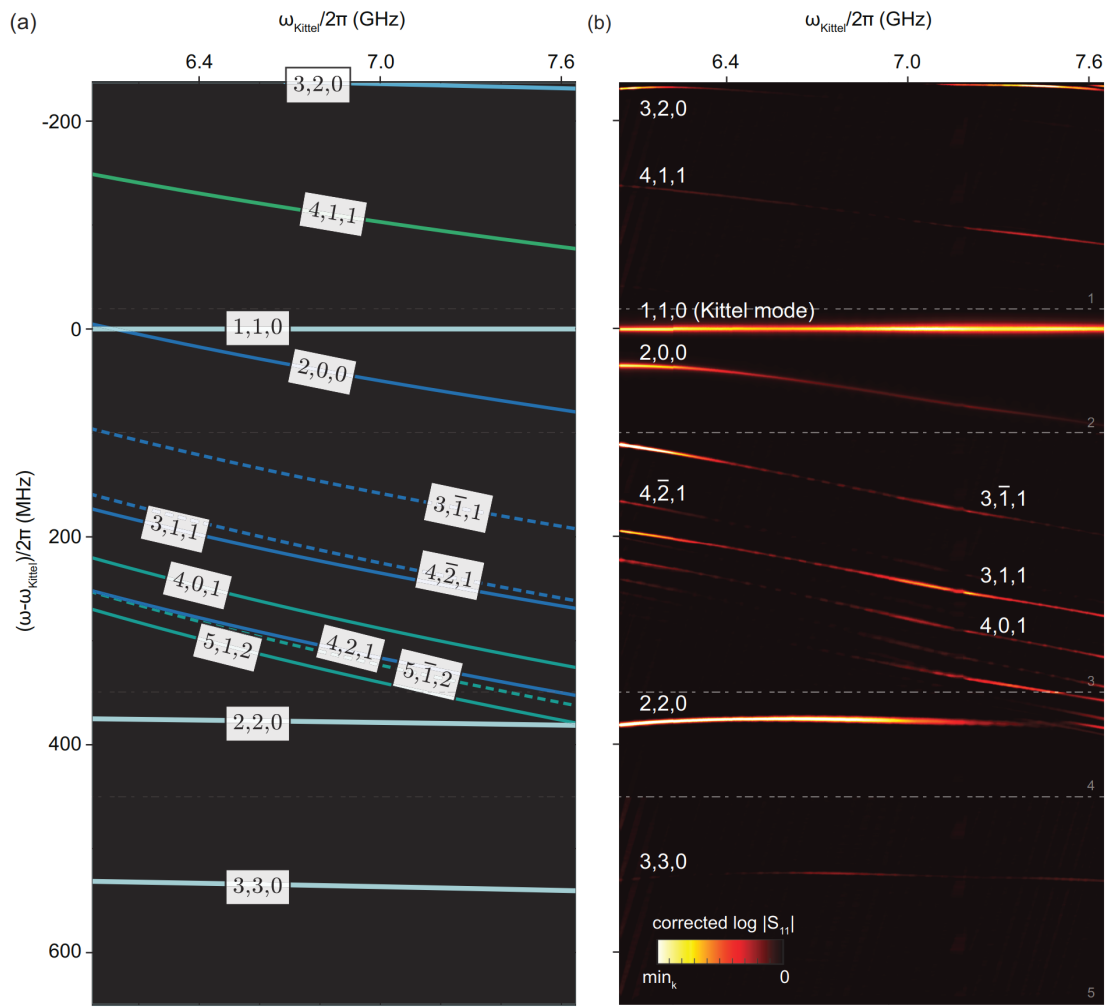}
\caption{\label{FigS1}(a) Theoretically predicted frequencies of the Walker modes and (b) experimentally observed FMR spectra for various values of applied magnetic field or equivalently the Kittel-mode frequency. In both plots, the vertical axis represents the frequency relative to the Kittel-mode frequency. In (b), $|S_{11}|$ is the normalized microwave reflection amplitude and the $\mathrm{min}_k$ represents the minimum values of $\log{|S_{11}|}$ in the region $k$ indicated in the figure.}
\end{figure*}

In this Supplemental Material, we describe the identification of the observed magnetostatic (Walker) modes based on the expected dependences of the mode frequencies on the applied static magnetic field.

The ferromagnetic sphere studied here has a diameter of $1\,$mm and its magnetization is saturated along the $[110]$ crystal axis by two identical permanent magnets in a magnetic circuit.  Reflected microwaves from a loop coil with the diameter of about $3\,$mm are analyzed by a network analyzer to exhibit ferromagnetic resonances (FMRs) of the yttrium iron garnet sphere.  The static magnetic field felt by the sphere can be tuned by a current injected into a coil integrated in the magnetic circuit.  Observed FMR spectra are composed of absorption dips due to the Walker modes and unwanted electrical interferences resulting from the cables, which can be omitted by the proper filtering.  

We investigate the dependence of the Walker-mode frequencies on the applied magnetic field by plotting the spectra with their frequencies referenced by the one of the Kittel mode. This procedure allows us to compare the theoretical values of the Walker-mode frequencies and the experimentally observed spectra, without the exact knowledge of the applied static magnetic field and of the internal field including the effects of the demagnetization due to the magneto-crystalline anisotropy.  The FMR spectra for various values of the static magnetic field, hence of the Kittel mode frequency, are plotted in Fig.~\ref{FigS1}(b).  The spectra are separately scaled in five regions (delimited by dashed lines) for visibility.  The theoretical values of the frequencies of Walker modes under various magnetic field strengths are plotted in Fig.~\ref{FigS1}(a). These theoretical values, originally calculated from the magnetostatic approximation, include corrections due to the propagation effect~\cite{Mercereau1959,FletcherPR1959, Plumier1962} and the magneto-crystalline anisotropy~\cite{Solt1960, Launets1971}. Additionally, we adopt an effective value for the saturation magnetization of 1940 G, 9\% above the typical tabulated value, which might be attributed to the inhomogeneity of the applied magnetic field. Given these, we can assign unambiguously indices for four modes, namely $(1,1,0), (3,\overline{1},1)$, $(3,1,1)$ and $(4,0,1)$.

\end{document}